\documentclass[12pt,preprint]{aastex} 

\newcommand{\kms}{$\rm km \, s^{-1}$}

\def\ks{km~s$^{-1}$\,}

\shorttitle{Broad H$\alpha$ wings in Young PNe} 
\shortauthors{Arrieta \& Torres-Peimbert} 

\begin{document} 

\title{Broad H$\alpha$ wings in Nebulae around Evolved Stars and 
in Young Planetary Nebulae} 

\author{\sc A. Arrieta \altaffilmark{1} 
and S. Torres-Peimbert %\altaffilmark{1} 
} 
\affil{Instituto de Astronom\'\i a (UNAM), 
Ap. Postal 70-264, M\'exico D.F. 04510 M\'exico} 
\email{anabel@astroscu.unam.mx} 

\altaffiltext{1} {Present address: Space Telescope Science Institute} 
\begin{abstract} 

Eleven objects that have been reported as proto-planetary nebula or as
young planetary nebulae that show very extended H$\alpha$ wings are
presented.  The extension of these wings is larger than 800 \ks. Data
for two symbiotic stars that show this same characteristic is also
presented. Raman scattering is the mechanism that best explains the
wings in 10 of the PNe and in the 2 symbiotic stars. In the PN
IRAS~20462+3416 the wing profile can be explained by very intense
stellar wind.

\end{abstract}


\vfill\eject 
\section{Introduction} 

As part of a study of proto-planetary nebulae and young planetary
nebulae we have observed 59 objects that have been selected for
fulfilling one or more of the following criteria: post-AGB stars with
evidence of high mass loss rate, or PNe with low degree of ionization
or with H$_2$, CO or OH molecular emission; all the objects observed
have IRAS colors typical of planetary nebulae. In the sample, 13
objects were outstanding for exhibiting extremely broad H$\alpha$
lines, of these, 10 have not been reported previously as having this
characteristic.  The H$\alpha$ profiles selected for being extremely
broadened are presented in this work.

Very wide H$\alpha$ emission lines have been reported previously in
other objects in the early stages of planetary nebulae phase; these
include seven post-AGB stars, five young PNe and some symbiotic stars
(Van de Steene, Wood \& van Hoof 2000; Lee 2000; Lee \& Hyung 2000;
Miranda, Torrelles \& Eiroa 1996; Balick 1989; L\'opez \& Meaburn
1983; Van Winckel, Duerbeck \& Schwarz 1993; Wallerstein 1978). Lee \&
Hyung (2000) advanced the proposal that the broad wings are produced
by Rayleigh-Raman scattering which involves atomic hydrogen, whereby
Ly$\alpha$ photons with a velocity width of a few $10^2$ \kms\, are
converted to optical photons and fill the H$\alpha$ broad region. In
the present work this mechanism is investigated further and alternate
possibilities for H$\alpha$ broadening are examined.

\section{Observations and reductions} 

The observations were carried out at the Observatorio Astron\'omico
Nacional in San Pedro M\'artir, Baja California with the 2.1-m
telescope and the REOSC echelle spectrograph (R $\sim $ 18,000 at
5,000 \AA) and a 1024$\times$1024 Tektronix detector that yields a
spectral resolution of 10.6 \ks and a spatial resolution of
0.99\arcsec\ per pixel. The 3600 to 6800\AA\ range was covered in 29
orders. For extended objects the slit was centered on the nucleus. The
observing log of those objects with wide H$\alpha$ wings is presented
in Table \ref{tlog}.

The data reduction was carried out with IRAF routines using `echelle'
and `ccdred' tasks.  Calibrated spectra for most of the objects were
obtained.  The orders were extracted using a 10 pixel window. A HeAr
lamp was used for wavelength calibration and the standard stars by
Hamuy et al. (1992) were observed for the flux calibration.

In several cases the objects exhibit extremely wide H$\alpha$ lines,
with full widths at zero intensity, FWZI, larger than 800 \kms.  In
order to ascertain whether the wide wings are real or due to
instrumental effects the ``number of counts at maximum'' vs. FWZI both
for the brightest unsaturated lines of the comparison lamp spectra and
for H$\alpha$ emission lines in our sample were compared.  Objects
with with FWZI $>$~1000 \kms\ and those with significantly broader
wings in H$\alpha$ than what can be considered instrumental were
selected for this study; namely, those with FWZI 3$\sigma$ above the
observed ones for comparable line intensities in the HeAr lamp.

\section{H$\alpha$ wings} 

\begin{figure} 
\centering 
\includegraphics[width=12cm]{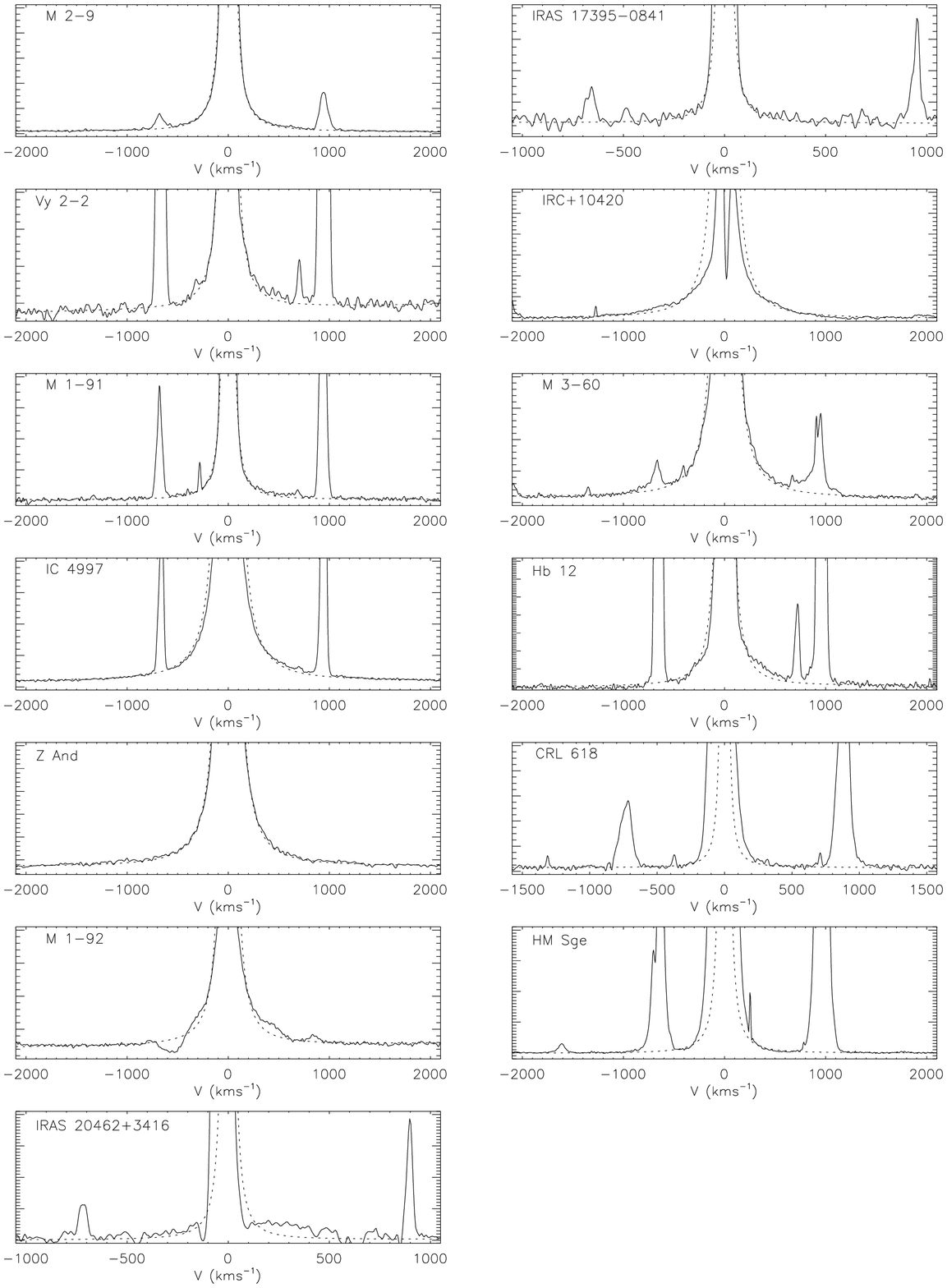} 
\caption{ 
H$\alpha$ \, profiles for the 13 selected objects of extreme
broadening.  In 12 cases a section of the spectrum between -2000 and
+2000 \kms is shown, and in one case -1000 and +1000 \kms.  Superposed
to the observations is the $1/v^2$ function that best adjusts to the
wing profiles within $\pm$~100~\kms of the core.
} 
\label{plotone} 
\end{figure}

The H$\alpha$ profiles for the 13 selected objects are shown on Figure
\ref{plotone}. To better visualize the line profiles, an enlargement
along the intensity to display the wings over the full H$\alpha$
profile is presented. In all cases the FWZI is significantly larger
than expected from a gaussian fit to the core of the profile.  No
correlation between the line intensity and the width of the wings was
found.

The main spectral characteristics of each object are presented in
Table \ref{tcarace}. It includes the ions present in emission and
absorption in our spectra and the stellar spectral type (as taken from
SIMBAD database).  For the 13 objects H$\alpha$ FWZI and H$\beta$ FWZI
where it is significantly larger than expected from a gaussian fit to
the core of the profile are presented in Table \ref{tcaracg}.  In this
table the most common classification in the literature for each
object, nebular morphology, possible binarity and type of H$\alpha$
line profile are included.

\section{Possible mechanisms for wing broadening} 

\subsection{Rotating disks} 

For this case maximum velocity would correspond to the circular
velocities at the surface of the exciting star, $$v_{max} = 437 \sqrt{
(m/M{_\odot})/(r/R{_\odot})}\,\,\, \rm{km \, s^{-1}\, \, ,} $$ which
for young PNe central stars of $m = 0.6-0.83$ M$_\odot$, and $r =
1-18$ R$_\odot$ would correspond to $v_{max} < 400 $ \kms, although
for symbiotic stars, where the hot component is small $r <~ 0.1$
R$_\odot$, larger velocities could be achieved. Nevertheless, for both
PNe and symbiotic stars, $v_{max}$ is not large enough to explain the
observed broadening which reaches up to 5000 \kms\ in some cases. Thus
the possibility that in PNe emission from a rotating disk would
explain by itself the extended wings can be ruled out.

\subsection{Stellar winds} 

The possibility for the line wings to be formed in the region
dominated by stellar winds was tested by examining the existing IUE
spectra for each object looking for evidence of P~Cyg profiles in
resonant lines of ions like \ion{C}{4}, \ion{C}{3}, \ion{He}{2},
\ion{Al}{3}, \ion{Mg}{2} and \ion{Si}{4}.

Only one object (IRAS~20462+3416) shows P~Cyg profile evidence both in
our optical spectra and in the UV. Furthermore, IRAS~20462+3416 shows
an anomalous broadened emission in H$\alpha$, that is +110 \kms\, from
the narrow emission line.  This profile shows significant deviations
from the $\Delta \lambda^{-2}$ profile that can be fit to the other
broad wings (see Figure \ref{plotone}).  The presence of P~Cyg
profiles both in optical and UV spectra, and the H$\alpha$ wing
profile suggests that stellar wind is the mechanism that is producing
these features.  In a separate paper (Arrieta, Torres-Peimbert,
Georgiev \& Koenigsberger 2003) it is shown that indeed the H$\alpha$
profile of this object can be explained by a very strong stellar wind.
The H$\alpha$ profiles for the rest of the objects cannot be explained
with stellar winds.

\subsection{Electron scattering} 

This mechanism has been proposed, and has been intensively studied, as
the line broadening mechanism in QSOs (e.g. Mathis 1970; Shields \&
McKee 1981; Lee 1999) and in WRs (e.g. Hillier 1991). Given that the
cross section of electron scattering is independent of wavelength it
is to be expected that other intense emission lines formed in the same
region as H$\alpha$ are similarly broad.

Extended wings in other bright emission lines of our optical spectra
were searched for. No broad extensions in forbidden lines in any of
the objects were found. The objects IRC+10420, M~1-92, HM~Sge, M~3-60,
and Z~And showed extended wings in H$\beta$. In all cases the width of
H$\beta$ is significantly smaller than that of H$\alpha$ (see Table
\ref{tcaracg}). It is possible that in the other objects the low
signal-to-noise ratio masks the low intensity extended wings. For
objects with broad H$\beta$ wings no correlations were found with the
presence of the H$\beta$ wings and the width of H$\alpha$, nor with
the H$\alpha$/H$\beta$ line ratios.

The necessary conditions for the H$\alpha$ wings to be produced by
electron scattering were investigated. Given that no significant
broadening was found in the forbidden lines, it is required for the
forbidden line region to be exterior to the electron scattering
region.  A two-region geometry was assumed: (a) a recombination line
emitting region where a substantial fraction of H$\alpha$ is produced,
dense enough to suppress the forbidden lines, and where most of the
scattering takes place ($N_e >~ 10^6$, $T_e \sim 10^4$) (b) and an
external low density forbidden line emitting region ($N_e =
10^4-10^6$~cm$^{-3}$, $T_e \sim 10^4$).  It is also possible to
consider that region (a) consists of a point source where most of the
H$\alpha$ is produced surrounded by a dense scattering region.

To derive the temperature in the electron scattering region from the
observed H$\alpha$ line profiles the treatment proposed by Mathis
(1970) was followed.  The main parameter is the width of the wings,
defined as $Y(2) \equiv F_{obs}(\Delta v_M)/F_{obs}(2\Delta v_M)$,
which essentially determines the temperature of the electron
scattering region, $T_{ES}$. The optical depth of the scattering
region, $\tau_{ES}$, can be derived from the ratio of the strength of
the unscattered core of the line relative to its total strength $w
\equiv F(unscattered)/F(total)$, which determines the optical depth of
the scattering region, $\tau_{ES}$ (Osterbrock 1980).  For the simple
case of a spherically symmetric scattering region of uniform density
and temperature for the case of M2-9 $Y(2) = 2.05$, and $w = 0.83$ can
be obtained.  From these values a temperature, $T_e \sim 10^8$ K, and
optical depth, $\tau_{ES} \sim 0.24$ can be found, the latter coupled
to a limiting size of $r_{ES} <~ 10^{12}$ cm, given by an unresolved
central core of 2\arcsec, yields a density $N_e >~ 10^{12}$ cm$^{-3}$
for the electron-scattering region.  From the values derived for
M~2-9, which is representative of the whole sample, it was considered
unlikely for this configuration to be the general case surrounding an
AGB star and thus to account for the extremely wide H$\alpha$ lines
observed in young PNe.

\subsection{Raman Scattering} 

Raman scattering describes the absorption of a photon, followed by the
immediate re-emission of another photon at different wavelength where
the intermediate state does not correspond to a true bound state of
the atom. Initially this process was suggested by Schmid (1989) as the
mechanism of production of the broad emission features at $\lambda
\lambda $6830 and 7088\AA\ found in 50\% of symbiotic stars; he suggested 
that the emission features are due to Raman scattering of the \ion{O}{6} 
resonance doublet $\lambda\lambda$ 1032 and 1038\AA\ by neutral hydrogen. 
Nussbaumer, Schmid \& Vogel (1989) proposed a list of uv lines of different 
ions (including \ion{S}{3}, \ion{He}{2}, \ion{O}{1}, \ion{O}{6} and 
\ion{C}{2}), with wavelengths close to Ly$\beta$, as candidates for Raman 
scattering on the ground state of neutral hydrogen; these Raman scattered 
lines would be expected to be found in the 6000 - 7000\AA\ , range. 
They noted the possibility that Raman scattering may also hold 
a clue to the broad line wings of H$\alpha$, occasionally observed in 
Seyfert galaxies and symbiotic stars.

Several of the emission features at different wavelengths that can be
attributed to Raman scattering have been identified in symbiotic stars
and in planetary nebulae. In symbiotic stars the identifications
include, among others: $\lambda\lambda$6830, 7088\AA\ from
\ion{O}{6} $\lambda\lambda$1032\AA\   and $\lambda\lambda$7021, 7052\AA\ 
from  \ion{C}{2} $\lambda\lambda$1036, 1037\AA\  in V1016 Cyg 
(Schmid 1989; Schild \& Schmid 1996) as well as $\lambda$4850\AA\ 
from \ion{He}{2} (2-8) $\lambda$972.1\AA, $\lambda$4331\AA\, from   
\ion{He}{2} (2-10)$\lambda$949.3\AA, and $\lambda$4975\AA\ from   
\ion{C}{3} $\lambda$977.0\AA\,  in RR Tel (van Groningen 1993). 
In planetary nebulae,  they include $\lambda$4850\AA\ in 
NGC 7027 (P\'equignot et al. 1997), and $\lambda$6545\AA\ from   
\ion{He}{2} (2-6) $\lambda$1025\AA\  in M2-9 (Lee et al. 2001).

Lee \& Hyung (2000) proposed that the broad H$\alpha$ wings of the PN
IC~4997 are formed through Raman scattering that involves atomic
hydrogen and, by which Ly$\beta$ photons with a velocity width of a
few 10$^2$ \kms\ are converted to optical photons and fill the
H$\alpha$ broad wing region. Their model fits the observations on the
blue wing from $v$ = 500 to 1500 \kms, it requires relatively strong
incident Ly$\beta$ flux from an unresolved core of high density ($N
\sim 10^9 - 10^{10}$ cm$^{-3}$) and a column density for the
scattering region of $n_{H^0} = 2 - 4 \times 10 ^{20}$ cm$^{-2}$. Lee
(2000) further proposed that the H$\alpha$ wings seen in symbiotic
stars can be fit to Raman scattered profiles. For the optically thin
case, where almost all the Ly$\beta$ photons are scattered not more
than once and assuming a flat incident Ly$\beta$ profile, to the first
order the wing profile can be approximated by a curve proportional to
$f(\Delta v) = \Delta v^{-2}$. Lee adjusted this curve to H$\alpha$
profile observations from 200 to 1000 \ks from the line center for 16
symbiotic stars.

Figure \ref{plotone} shows the fit of curves proportional to $\Delta
v^{-2}$ to the wings of the 13 objects under consideration in this
paper.  In general the fits are satisfactory for 12 objects, excepting
IRAS 20462+3416 where it does not at all fit the observed profile; the
curves match the observations starting at 200 - 300 \kms\ from the
center of the line, to the outermost regions where the signal becomes
too faint to be significant, this extreme velocity reaches up to 500 -
1500 \kms.  The differences in some objects can be explained in terms
of non uniform motions in the ionized regions.  The kinematics of
these cases will be analyzed in later studies.  In particular, CRL 618
and HM Sge show wider cores than the expected profile, while IRC+10420
shows a narrower core; M1-92 shows significant differences in the 300
- 800 \ks\ region, which have been interpreted as a jet of material
along the line of sight (Arrieta, Torres-Peimbert, \& Georgiev
2000). The case of IRAS~20462+3416 is clearly different, because it
does not fit at all such a $f(\Delta v) = \Delta v^{-2}$ profile,
neither in the core, nor in the wings.  As mentioned in the previous
section the interpretation of this profile corresponds to a very large
mass loss rate (Arrieta, Torres-Peimbert, Georgiev \& Koenigsberger
2003) .

The width required for Ly$\beta$ emission to be responsible for the
H$\alpha$ wings has been investigated in more detail. Since the width
of the scattered H$\alpha$ is proportional to the initial width of the
Ly$\beta$ line, then, $\Delta v_{Ly\beta} = \Delta v_{H\alpha}/6.4$ is
expected.  An estimate of the Ly$\beta$ width can be given by other UV
emission lines widths.  A search in the UV high resolution spectra of
MAST (Multimission Archive at Space Telescope) was carried out for the
objects under consideration where data for Z~And, HM~Sge and IC~4997
were found. In the cases of Z~And and HM~Sge the \ion{Si}{3}]
$\lambda$1892\AA\ line has width of $\sim 500$ \ks, consistent with
the $\Delta v_{Ly\beta}$ width ($\sim 625$ and $\sim 468$ \ks)
required to produce the observed broadened H$\alpha$ lines (as given
in Table \ref{tcaracg}).  In the case of IC~4997 the width of the
\ion{Si}{3}] $\lambda$1892\AA\ line is of only $\sim 160$ \ks.
Although Lee et al. (2001) consider that since the Ly$\beta$
underlying continuum emission is not likely to show a flat profile, it
may be more plausible that the broad H$\alpha$ wings are formed by
Raman scattering of the continuum photons around Ly$\beta$.

Evidence of additional Raman scattered features in our optical spectra
was searched for. The $\lambda$6830\AA\ feature was found to be
present in Z~And and weakly in HM~Sge with a FWZI $\sim$1500 \kms\
(this feature has been studied previously by Birriel et al. 1998;
Schmid et al. 2000).  Features around this wavelength were found in
M~2-9, IRAS~17395-0841, IRC+10420, M~3-60, IC~4997 and possibly in
M~1-92 with FWZI around a few 10$^2$ \ks \, and not a few 10$^3$ \ks
\, as is the case in symbiotic stars. The $\lambda$6545\AA\ feature in
M~2-9 and M~1-92 as well as $\lambda$4851\AA\ in IC~4997 with FWZI
around few 10$^2$ of \ks were also found. In that same sense, the
H$\beta$ wings found in IRC+10420, M~1-91, HM~Sge, M3-60, Hb~12 and
Z~And (see Table 3) can be explained by Raman scattering of the
Ly$\gamma$ line by the neutral hydrogen component.

In order to obtain additional indicators of the presence of neutral or
molecular material along the line of sight a search in the literature
was carried out for data on 21-cm measurements and the column density
of atomic hydrogen.  Only in the symbiotic star HM~Sge and the
planetary nebula IC~4997 21-cm column densities have been measured; in
both cases values of $n_{H^0} \sim 4 \times 10 ^{20}$ cm$^{-2}$ were
determined.  Also, following Dinerstein et al. (1995), evidence of
nebular \ion{Na}{1} doublet $\lambda\lambda$5889, 5895\AA \, from our
optical spectra was looked for, where \ion{Na}{1} nebular components
in 11 of our 13 objects were found; those with bright optical
continuum and favorable radial velocities, (objects with radial
velocities well separated from those expected for the interstellar
material) which allow the separation of the nebular component and the
narrow interstellar one.  In addition a search in the literature for
CO rotational transitions in radio wavelengths, roto-vibrational ones
in the near infrared and H$_2$ roto-vibrational transitions was
performed.  The results are given in Table \ref{tneutra} where
evidence of a neutral component at least in one of the indicators is
listed.

\section{Discussion and summary } 

We confirm that Raman scattering is the most probable mechanism for
the formation of the H$\alpha$ broad wings in 12 of the 13 objects
under consideration.  This is supported by a fit of the profile to a
$\Delta v^{-2}$ law, by the presence of Raman produced features in
emission, and by additional indicators of the presence of a
significant neutral hydrogen component.

Evidence of neutral components in other 29 objects observed with
H$\alpha$ in emission was searched for, in order to compare with the
sample presented here.  However there was not enough information in
the literature to derive significant statistical data between the
objects that show extreme wing widening and those that do not show it.
Only in the case of the hydrogen molecule it was found that those
objects with broad H$\alpha$ wings have H$_2$, while those that are
not broadened, do not.

Most of the objects with broad lines seem to have common
characteristics: bipolar morphology, and composite emission line
profiles, furthermore, they are compact objects in the process of
forming a planetary nebula and very probably they are sites of wind
interaction.

More detailed studies of the conditions for Raman scattering to take
place and thus for the presence of surrounding neutral hydrogen could
give us information about the processes that take place in giving rise
to bipolar planetary nebulae.

\acknowledgements 
We are grateful to M. Bautista, L. Georgiev y E. Villaver for fruitful
discussions and to H. M. Schmid for many important suggestions to the
original manuscript. Support from DGAPA-IN100799, DGAPA-IN114601 and
CONACyT-25451E grants is acknowledged.

\clearpage 

\clearpage 

\begin{table} 
\begin{center} 

\caption[]{Observing log of objects with extended wings } 
\label{tlog}
\begin{tabular}{lllr} 
\tableline\tableline 

        &     &  date    &  exposure  \\ 
Object         & PNG & observed &  time (sec)   \\ 
\tableline 
CRL 618          &  166.4-06.5  &  5 Sep 95 & 1800  \\ 
M 2-9            &  010.8+18.0   &  23 Apr 96 &  600  \\ 
IRAS 17395-0841  &  017.0+11.1   &   10 Jun 97 & 1500   \\ 
Vy 2-2           &  045.4-02.7   &   3 Jun 97  &  900   \\ 
IRC+10420        &    -           &   27 Sep 95 &  900   \\   
M 1-91           &  061.3+03.6   &   11 Jun 97 & 2700    \\ 
M 1-92           &    -           &   10 Jun 97 &  300     \\ 
HM Sge           &    -           &   23 Apr 96 & 600   \\ 
M 3-60           &    -           &   27 Sep 95 &  900    \\ 
IC 4997          &  058.3-10.9   &   9 Jun 97 &  180    \\ 
IRAS 20462+3416  &    -           &   6 Sep 97 & 1600   \\ 
Hb 12            &  111.8-02.8   &   1 Sep 97 & 600   \\ 
Z And            &    -           &   25 Sep 95 & 900 \\ 
\tableline 
\end{tabular} 
\end{center} 
\end{table}

\clearpage 

%\begin{table} 
%\begin{center} 
\begin{deluxetable}{lllll} 
\small 
\tablewidth{0pt} 
\tablecaption{Nebular and stellar spectral characteristics of the objects 
with very extended H$\alpha$ Wings. }
\label{tcarace} 
%\begin{tabular}{lllll} 
%\tableline\tableline 
\tablehead{ 
\colhead{Object}          &   \colhead{Emission lines}                      & 
\colhead{Absorption lines }   & \colhead{Sp. type}} 
\startdata 
CRL 618         & \ion{H}{1},[\ion{Fe}{2}],[\ion{N}{2}],[\ion{O}{1}],[\ion{O}{3}],[\ion{S}{2}]            &  non detectable
      & B0            \\ 
M 2-9           & \ion{H}{1},\ion{He}{1},\ion{He}{2},[\ion{Ar}{3}],[\ion{Cr}{2}],[\ion{Fe}{1}],          &  non detectable
      & Be            \\ 
                &[\ion{Fe}{2}],[\ion{Fe}{3}],[\ion{Fe}{4}],[\ion{N}{2}],[\ion{Ni}{2}],
                                                  \\ 
                &[\ion{Ni}{3}],[\ion{Ni}{4}],[\ion{O}{1}],[\ion{O}{2}],[\ion{O}{3}],      &                       &
              \\ 
                & [\ion{S}{1}],[\ion{S}{2}],\ion{Si}{2}                                   &                       &
              \\   
IRAS 17395-0841 & \ion{H}{1},\ion{He}{1},[\ion{N}{2}],[\ion{Ne}{3}],[\ion{O}{1}],[\ion{O}{3}],[\ion{S}{2}]       &  non
detectable       & --             \\ 
Vy 2-2          & \ion{H}{1},\ion{He}{2},[\ion{Ar}{3}],[\ion{N}{2}],[\ion{Ne}{3}],[\ion{O}{1}],[\ion{O}{2}],    &  non detectable
      & Of            \\ 
                & [\ion{O}{3}],[\ion{S}{2}],[\ion{S}{3}]                          &                       &               \\ 
IRC 10420       & \ion{H}{1},\ion{Cr}{2},\ion{Fe}{1},\ion{Fe}{2},\ion{Mg}{1},\ion{Sc}{2},\ion{Sr}{2},\ion{Ti}{2}          &
\ion{H}{1},\ion{Ca}{2},\ion{Cr}{2},\ion{Fe}{1}, & F8 Ia         \\ 
                &                                              & \ion{Fe}{2},\ion{Na}{1},\ion{Si}{2},\ion{Ti}{2}
                        \\ 
M1-91           & \ion{H}{1},\ion{He}{1},[\ion{Fe}{2}],[\ion{N}{2}],[\ion{O}{1}],[\ion{O}{3}], &  non detectable       & Be
        \\ 
                & [\ion{S}{2}],[\ion{S}{3}]         & &  \\ 
M 1-92          & \ion{H}{1},\ion{He}{1},[\ion{Ca}{2}],[\ion{Cr}{2}],[\ion{Fe}{2}],[\ion{N}{1}],  &
 \ion{H}{1},\ion{Ca}{2},\ion{Fe}{2},                    & B0.5IV   \\ 
                & [\ion{N}{2}],[\ion{O}{1}],[\ion{O}{3}],[\ion{S}{2}],\ion{Ti}{2}                           &
 \ion{He}{1},\ion{N}{1}                     &               \\ 
HM Sge          & \ion{H}{1},\ion{He}{1},\ion{He}{2},[\ion{Ar}{4}],\ion{C}{2},\ion{C}{3},[\ion{Fe}{2}],   & non detectable
       &  M            \\ 
                & [\ion{Fe}{3}],[\ion{Fe}{4}],[\ion{Fe}{5}],[\ion{Fe}{7}],[\ion{K}{4}],\ion{Mg}{1},   &                       &
             \\ 
                & [\ion{N}{2}],\ion{N}{2}I,\ion{Ne}{2},[\ion{Ne}{3}],\ion{O}{1},\ion{O}{2},\ion{O}{3},     &  &
                              \\ 
                & [\ion{S}{2}],[\ion{S}{3}],\ion{Si}{2} & & \\ 
M 3-60          & \ion{H}{1},\ion{Fe}{2},[\ion{N}{2}],[\ion{O}{1}],[\ion{S}{2}]                     & non detectable        &  B
           \\ 
IC 4997         & \ion{H}{1},\ion{He}{1},[\ion{Fe}{2}],[\ion{N}{2}],[\ion{O}{1}],[\ion{O}{3}],[\ion{S}{2}]        &  non
detectable       &               \\ 
IRAS 20462+3416  & \ion{H}{1},[\ion{N}{2}],[\ion{S}{2}],\ion{Si}{2}                           &
\ion{H}{1},\ion{He}{1},\ion{He}{2},\ion{Al}{3}, &  B           \\ 
                &                                              & \ion{Ca}{2},\ion{N}{2},\ion{Ne}{1},\ion{O}{2},
                 \\ 
& & \ion{S}{2},\ion{Si}{2} \\ 
Hb 12           & \ion{H}{1},\ion{He}{1},\ion{He}{2},\ion{Ar}{4},\ion{C}{2},\ion{Cl}{2},\ion{Cl}{3},[\ion{Fe}{2}],       &  non
detectable       &   --           \\ 
                & [\ion{Fe}{3}],\ion{Mg}{1},[\ion{N}{2}],\ion{N}{3},[\ion{Ne}{3}],[\ion{O}{1}],[\ion{O}{2}],    &
                      &               \\ 
                & [\ion{O}{3}],[\ion{S}{2}],[\ion{S}{3}],\ion{Si}{2},\ion{Si}{3}               &                       &
              \\ 
Z And           & \ion{H}{1}, \ion{He}{1}, \ion{He}{2}, [\ion{Cr}{1}], \ion{Fe}{2}, [\ion{Fe}{7}],          &  TiO
                & M6.5          \\ 
                & [\ion{K}{5}], [\ion{Mn}{2}], [\ion{O}{1}],\ion{Si}{2}                      &                      &
                 \\ 
\enddata 
\end{deluxetable} 
%\end{center} 
%\end{table} 

\clearpage 
\begin{table} 
\begin{center} 

\caption[]{Characteristics of the objects with very wide  H$\alpha$ wings (classification, morphology, 
binarity, profile type and  widths of  H$\alpha$  and H$\beta$). } 
\label{tcaracg}
\begin{tabular}{lllllrr} 
\tableline\tableline 
         &                   &             &          &         & H$\alpha$  &  H$\beta$   \\ 
         &                   &             &          &         & FWZI       & FWZI        \\ 
 Object  &   Classif.        &   Morphology & Binarity  & Profile & (\kms) &  (\kms)          \\ 
\tableline 
CRL 618          & Proto-PN & bipolar      &  -       & double    & 2300  &  -              \\ 
M 2-9            & YPN      & bipolar      & probable (1) & double    & 5000  &  -              \\ 
IRAS 17395-0841  & Proto-PN & non resolved  &  -       & simple    & 800   &  -              \\ 
Vy 2-2           & YPN      & bipolar      &  -      & simple    & 1400  &  -              \\ 
IRC+10420        & OH/IR    & bipolar      &  no (2)      & comp.    & 2600  &  1750           \\ 
M 1-91           & YPN      & bipolar      & probable (3) & double    & 1100  &  -              \\ 
M 1-92           & YPN      & bipolar      &  yes (4)      & double    & 2900  &  600  \\ 
HM Sge           & symb.    & bipolar      &  yes (5)      & double    & 3000  & 1300  \\ 
M 3-60           & YPN      & non resolved  &  -           & double    & 2400  &  800  \\ 
IC 4997          & YPN      & bipolar      &  -           & double    & 5100  &  -    \\ 
IRAS 20462+3416  & YPN      & oblate       &  -           & PCyg     & 2200  &  -    \\ 
Hb 12            & YPN      & bipolar      &  -           & simple    & 1800  &  -    \\ 
Z And            & symb.    & non resolved  &  yes (6)  & double    & 4000  &  1000 \\ 
\tableline 
\end{tabular} 
\tablerefs{(1) Schwarz et al. 1997; 
(2) Hrivnak et al. 1989; 
(3) Rodr\'{\i}guez et al. 2001; 
(4) Feibelman \& Bruhweiler 1990; 
(5) Taranova \& Yudin 1983; 
(6) Mikolajewska \& Kenyon 1996} 

\end{center} 
\end{table} 

\clearpage 

\begin{table} 
\begin{center} 
\caption{\small Evidence of neutral components along the line of 
sight of those objects with wide H$\alpha$ lines.   } 
\label{tneutra}
{\small 
\begin{tabular}{lclll} 
\tableline \tableline 
       & H I          &              &    &   \\ 
Object           & (10$^{20}{\rm cm}^{-2}$) & \ion{Na}{1} &    H$_2$      &   CO  \\ 
\tableline 
CRL 618          & -           &    yes                          &  yes (3)   &  yes (5) \\ 
M 2-9            & -           &    yes                          &  yes (3)   &  yes (5) \\ 
IRAS 17395-0841  & -                      &  yes               &  -        &   -  \\ 
Vy 2-2           & -                      &  yes (2)           &  yes (3)   &  no (5) \\ 
IRC+10420        & -                     &  yes               &  -        &  yes (6)  \\ 
M 1-91           & -                      &  no               &  yes (4)   &  no (7) \\ 
M 1-92           & -                     &  yes               &  yes (3)   &  yes (8) \\ 
HM Sge           & 4 (1)                  &  no               &  -        &  yes (9) \\ 
M 3-60           & -                       &  yes               &  -        &   -  \\ 
IC 4997          & 3.8 (2)               &  yes (2)           &  no (4)   &  no (5) \\ 
IRAS 20462+3416  & -                       &  yes               &  -        &  -   \\ 
Hb 12            & -                      &  yes               &  yes (3)   &  no (2) \\ 
Z And            & -                     &  yes               &  -        &  -   \\ 
\tableline \tableline 
\end{tabular} 
} 
\end{center} 
\tablerefs{ (1) Leahy et al. 1990; 
(2) Dinerstein et al. 1995; 
(3) Hora et al. 1999; 
(4) Kastner et al. 1996; 
(5) Huggins \& Healy 1989; 
(6) Knapp \& Morris 1985; 
(7) Josselin et al. 2000; 
(8) Alcolea et al. 2000; 
(9) Mueller \& Nussbaumer 1985.} 
\end{table} 

\end{document}